\documentclass[twocolumn,floatfix,citeautoscript,nofootinbib,superscriptaddress]{revtex4-2}
\usepackage{eurosym}
\usepackage{amsfonts}
\usepackage{mathrsfs}
\usepackage{xspace}
\usepackage{subfigure}
\usepackage{longtable}
\usepackage{xspace}
\usepackage{multirow}
\usepackage{tabularx}
\usepackage{amsmath}
\usepackage{amssymb}
\usepackage{graphicx}
\usepackage{dcolumn}
\usepackage{color}
\usepackage{bm}
\usepackage[colorlinks,urlcolor=blue,citecolor=blue,linkcolor=blue]{hyperref}
\usepackage[normalem]{ulem}
\usepackage{float}

\begin{document}

\title{Observation of dense collisional soliton complexes\\ in a two-component Bose-Einstein condensate}
\author{Sean M. Mossman}
\affiliation{Department of Physics and Biophysics, University of San Diego, San Diego, CA 92110, United States}
\affiliation{Department of Physics and Astronomy, Washington State University, Pullman, Washington 99164-2814, United States}
\thanks{Email address: smossman@sandiego.edu}
\author{Garyfallia C. Katsimiga}
\affiliation{Department of Physics, Missouri University of Science and Technology, Rolla, MO 65409, United States}
\affiliation{Department of Mathematics and Statistics, University of Massachusetts, Amherst MA 01003-4515, United States}
\author{Simeon I. Mistakidis}
\affiliation{Department of Physics, Missouri University of Science and Technology, Rolla, MO 65409, United States}
\author{Alejandro Romero-Ros}
\affiliation{Center for Optical Quantum Technologies, Department of Physics,
University of Hamburg, Luruper Chaussee 149, 22761 Hamburg, Germany}
\author{Thomas M. Bersano}
\affiliation{Department of Physics and Astronomy, Washington State University, Pullman, Washington 99164-2814, United States}
\author{Peter Schmelcher}
\affiliation{Center for Optical Quantum Technologies, Department of Physics,
University of Hamburg, Luruper Chaussee 149, 22761 Hamburg, Germany}
\affiliation{The Hamburg Centre for Ultrafast Imaging,
University of Hamburg, Luruper Chaussee 149, 22761 Hamburg,
Germany}
\author{Panayotis G. Kevrekidis}
\affiliation{Department of Mathematics and Statistics, University of Massachusetts, Amherst MA 01003-4515, United States}
\author{Peter Engels}
\affiliation{Department of Physics and Astronomy, Washington State University, Pullman, Washington 99164-2814, United States}

\begin{abstract}
\noindent\large\textbf{Abstract} 

Solitons are nonlinear solitary waves which maintain their shape over time and through collisions, occurring in a variety of nonlinear media from plasmas to optics.
We present an experimental and theoretical study of hydrodynamic phenomena in a two-component atomic Bose-Einstein condensate where a soliton array emerges from the imprinting of a periodic spin pattern by a microwave pulse-based winding technique.
We observe the ensuing dynamics which include shape deformations, the emergence of dark-antidark solitons, apparent spatial frequency tripling, and decay and revival of contrast related to soliton collisions. 
For the densest arrays, we obtain soliton complexes where solitons undergo continued collisions for long evolution times providing an avenue towards the investigation of soliton gases in atomic condensates. 
\end{abstract}

\maketitle

{\vspace{0.25cm}\noindent\large\textbf{Introduction}}

Since their first observation in water waves~\cite{russell1885wave,korteweg1895waves}, the dynamics of solitary wave structures has evolved into a major thrust within nonlinear science.
These dispersionless, localized, coherent structures, which can undergo collisions without changing shape, are found in a wide range of integrable and near-integrable systems with broad applicability in optics~\cite{Kivshar2003}, atomic physics~\cite{siambook}, plasmas~\cite{kono}, fluids~\cite{ablowitz2} and other fields~\cite{ablowitz2,infeld}.

\begin{figure}[ht]
\centering
\includegraphics[width=\columnwidth]{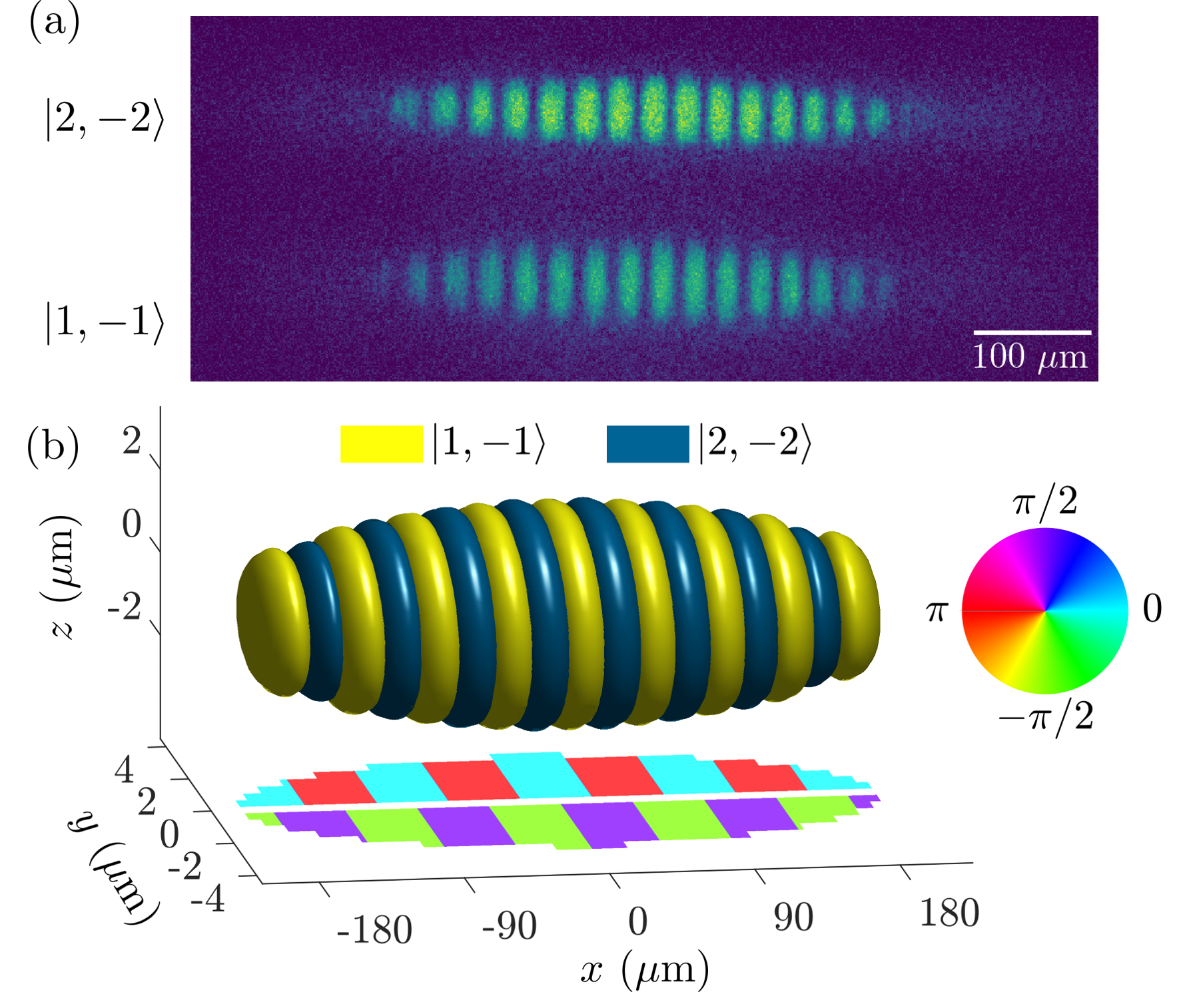} 
\caption{Initial state of the wound two-component BEC. (a) Example absorption image of the wound configuration after a winding time of $\tau=40$~ms with the $\lvert 2,-2\rangle$ (top) imaged after 6~ms time-of-flight and  $\lvert 1,-1\rangle$ (bottom) states imaged after 7.5~ms time-of-flight.
(b) Example of a numerical initial configuration of the two-component elongated BEC in situ. On the $x-y$ plane the condensate phase pattern corresponding to the $\lvert 2,-2\rangle$ ($\lvert 1,-1\rangle$) state is projected along the negative (positive) $y-$axis.}
\label{fig:winding_intro}
\end{figure}

Dilute gas Bose-Einstein condensates (BECs)~\cite{pethick,becbook2} offer a highly flexible and controllable platform towards investigating the nonlinear dynamics and interactions of such solitary wave structures~\cite{siambook}.
The experimental realization of multi-component BECs, discussed in Refs.~\cite{kawaguchi2012spinor,stamper2013spinor}, has led to an additional wealth of nonlinear states including dark-bright (DB), dark-dark, and dark-antidark vector solitons, among many others summarized, e.g., in Ref.~\cite{kevrekidis2016solitons}, as well as three-component \cite{Bersano2018,lannig2020collisions} and magnetic solitons~\cite{PhysRevLett.116.160402}.
While these works merely represent a small fraction of recent experimental and theoretical developments, they exemplify the remarkable flexibility offered by ultracold atomic systems for engineering and interrogating superfluid hydrodynamics.

Despite the intense research efforts directed towards solitons and their dynamics, most of the associated experimental BEC studies have concentrated on individual solitons or very small clusters (or molecules) thereof~\cite{Katsimiga2020} and their  interactions~\cite{lannig2020collisions}.
However, over the past few years there has been a substantial interest devoted to the realization and exploration of soliton gases, given their intriguing generalized hydrodynamic properties~\cite{PhysRevLett.120.045301}. 
First theoretically introduced in 1971 as a dilute soliton gas~\cite{zakharov1971}, the concept was later extended to the dense soliton gas in the theory works of Refs.~\cite{El2003,El2005,El2020}. 
However, experimental evidence for the realization of a soliton gas has been obtained only recently in the setting of shallow water waves~\cite{PhysRevLett.122.214502,Suret2020}. 
The relevant theory for integrable systems has been summarized in Ref.~\cite{El_2021}. 
These investigations, along with earlier efforts, namely on light pulses in optical fiber ring resonators~\cite{PhysRevE.55.7720} and also ones in soliton turbulence in shallow water waves~\cite{PhysRevLett.113.108501}, has changed our perspective on soliton lattices, soliton fluids, and soliton gases, as well as transitions between them~\cite{PhysRevE.91.032905}. 
Each of these experimental platforms provides a different framework for probing the physics underlying nonlinear hydrodynamics, each with their own advantages and limitations. In this work we will make use of the tunability, reproducibility, and mature numerical modeling methods of ultracold atomic gases to provide a different perspective on these topics. 

We present here a combined experimental and numerical study of hydrodynamic excitations arising from a periodic phase winding in a two-component spinor BEC.
Previously, experimental efforts along these lines have led to different examples of pattern formation in atomic BECs including Faraday waves~\cite{PhysRevLett.98.095301, kwon2021spontaneous, HernandezRajkov_2021, PhysRevLett.128.210401}, space-time crystals~\cite{PhysRevLett.121.185301}, and bright soliton trains produced from a dynamical (modulational) instability ~\cite{strecker,PhysRevA.96.041601,doi:10.1126/science.aal3220}.
In contrast to these earlier works, our method provides an initial condition from which exceptionally regular, highly tunable solitonic arrays will develop. 

We experimentally produce the initial state through a Ramsey pulse sequence in the presence of a small magnetic field gradient which produces an alternating magnetization pattern as shown in Fig.~\ref{fig:winding_intro}.
The periodicity of the pattern depends on the winding time $\tau$ between the two Ramsey pulses, with more time producing a finer pattern.
The experimental observations are made through state-selective absorption imaging as shown in Fig.~\ref{fig:winding_intro}(a).
Complementary numerical simulations are produced by solving the nonlinear Schr\"odinger equation in three dimensions, using a phase wound initial state as shown in Fig.~\ref{fig:winding_intro}(b).
This sinusoidal magnetization pattern then evolves under nonlinear interactions into different hydrodynamic phenomena depending on the spacing of the initial winding pattern.
In our exposition of the observed dynamics, we demonstrate how a broad magnetization pattern, under the influence of a magnetic gradient, leads to the emergence of solitons from an array of hydrodynamic shock fronts.  
We then study finer initial windings where we observe the formation of an array of dark-anitdark solitons which subsequently interact and evolve.
For our tightest arrays, we obtain dense, long-lived collisional soliton complexes spanning essentially the full extent of the BEC. 

{\vspace{0.25cm}\noindent\large\textbf{Results}}

\begin{figure}[t]
    \centering
    \includegraphics{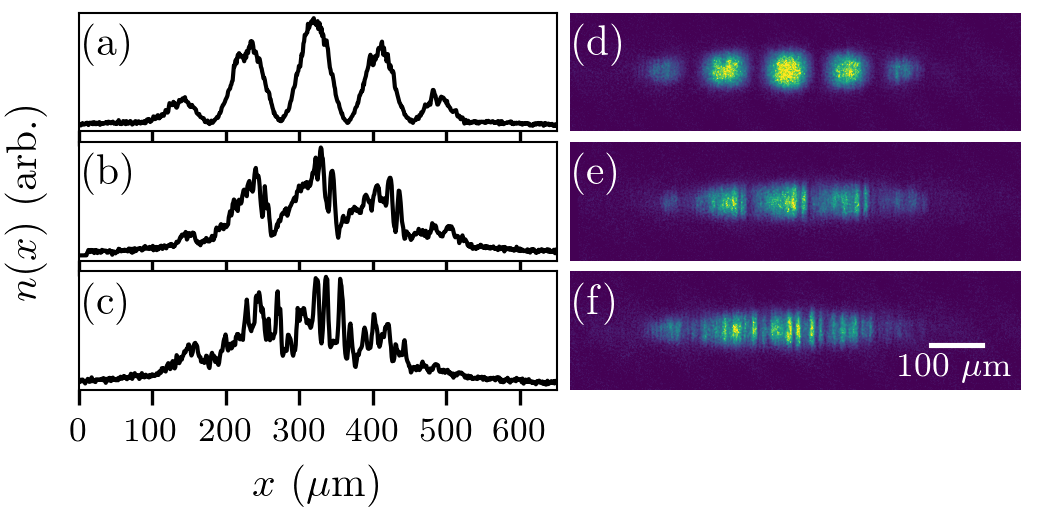}
    \caption{Shock-wave formation and gradient catastrophe. Experimental images of the 
    dynamics after a winding time of 10 ms in the presence of slight counterflow. Evolution times after the end of winding are (a, d) 10~ms, (b, e) 110~ms, and (c, f) 130~ms. Panels (a-c) show integrated cross sections of panels (d-f).
	}
    \label{fig:10ms_winding}
\end{figure}

We begin discussing the dynamics in mixtures with a wide initial winding pattern as shown in Fig.~\ref{fig:10ms_winding}(a), which corresponds to an experimental winding time of 10~ms. 
In these and all following experimental images, only one of the two spin states is shown as the second component forms the complementary pattern producing a nearly unmodulated total density (See Supplementary Note 1).
While later in this work we remove the axial magnetic gradient after the winding process to observe the unbiased evolution, for the data shown in Fig.~\ref{fig:10ms_winding} we maintain the magnetic gradient of 5~mG/cm along the long axis of the BEC to induce a small amount of counterflow between the two components. 
During the in-trap evolution, the initial sinusoidal shapes of the windings, seen in  Fig.~\ref{fig:10ms_winding}(a, d), steepen on their right side, approaching a gradient catastrophe~\cite{el2016}.
The steepening is arrested by the formation of dark notches which first emerge at the steep edges [Fig.~\ref{fig:10ms_winding}(b, e)] and then spread throughout the cloud [Fig.~\ref{fig:10ms_winding}(c, f)], similarly to previous work in superfluid counterflow~\cite{hamner2011}.  
Such dynamics are a hallmark phenomenon of dispersive hydrodynamics where, in the absence of viscosity, a gradient catastrophe is regularized by the formation of dispersive shock waves~\cite{el2016}. 
The shock structures involve the formation of solitons, which persist for long times. 
These observations lay the foundation for the following dynamics in more tightly wound clouds, where the size of each wound domain offers less space for the shock-like dynamics to unfold.

\begin{figure}[t]
\centering
\includegraphics[width=\columnwidth]{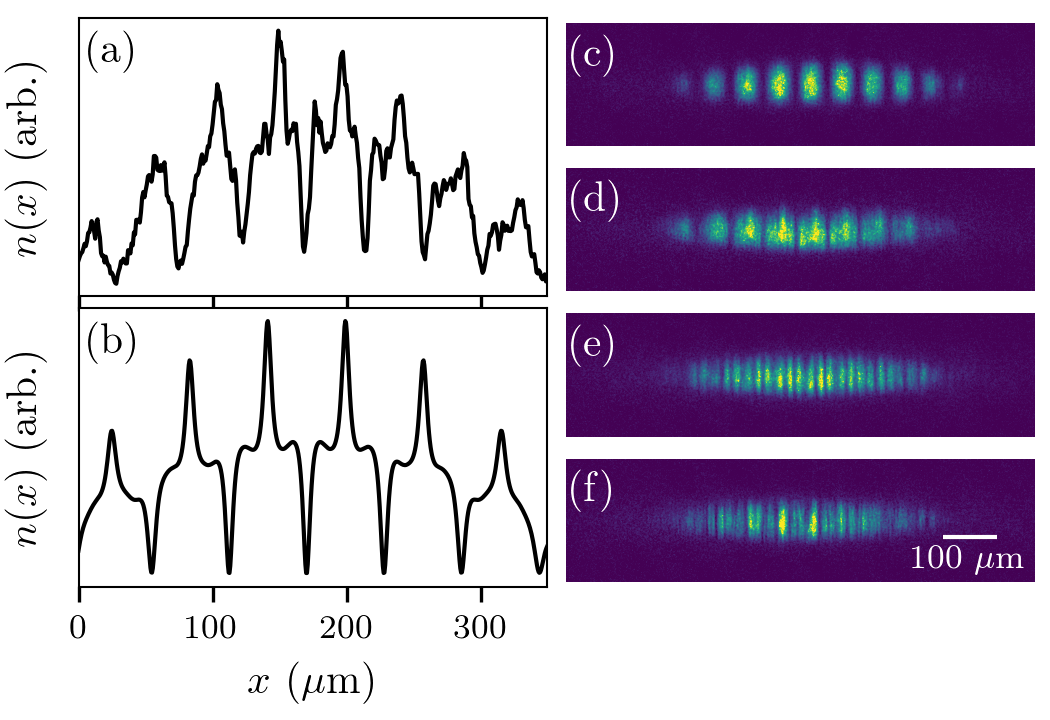} \caption{Emergence of antidark solitons after a winding time of 20 ms. (a) An integrated cross section of after 50 ms of evolution time reveals the formation of plateaus with antidark solitons, and (b) the corresponding cross section of 3D simulations shows similar emergence of antidark solitons. Absorption images show dynamics for evolution times after the winding process for (c) 0 ms, (d) 50 ms, (e) 80 ms and (f) 135 ms. }
\label{fig:20ms_winding}
\end{figure}

Next, we consider a case where the initial pattern size is too small for a full dispersive shock wave train to evolve, but wide enough for the initial stages of such dynamics to still emerge. 
This case, showcased in Fig.~\ref{fig:20ms_winding}, is reached after 20~ms winding (resulting in an initial pattern periodicity of 47~$\mu$m).
Here, no magnetic gradient is applied during the evolution following the initial winding, and dynamics set in symmetrically on either side of each winding. 
The sinusoidal shape of the initial windings begins by deforming and becoming triangular [Fig.~\ref{fig:20ms_winding}(c)]. 
This evolution transitions to the formation of plateaus with a high-density peak in the center of each plateau as depicted in Fig.~\ref{fig:20ms_winding} (d), taken after an evolution time of 50 ms. 
The corresponding density profile is shown in Fig.~\ref{fig:20ms_winding}(a) and is compared against the prediction of the 3D Gross-Pitaevskii equation for this winding configuration, shown in Fig.~\ref{fig:20ms_winding}(b).
These peaks, corroborated through our simulations, have the character of antidark solitons: bright solitonic peaks sitting on a finite background with a corresponding dark soliton in the other spin component~\cite{danaila2016vector,Katsimiga2020}. 
On both sides of each antidark peak, the steep edges of the plateaus lead to further evolution, as intuitively expected from the edge dynamics seen in Fig.~\ref{fig:10ms_winding} (or, mathematically, from the well-known Riemann problem of nonlinear differential equations~\cite{el2016}).
Notches appear and deepen, leading in this case to a tripling of the initial spatial period [Fig.~\ref{fig:20ms_winding}(e) after 80~ms]. 
The width of these features is approximately on the length scale that we experimentally observe for DB solitons in our system, as demonstrated in Fig.~\ref{fig:10ms_winding}, restricting the space against further soliton formation. 
Instead, the pattern becomes more irregular in amplitude at longer evolution times [Fig.~\ref{fig:20ms_winding}(f) at 135~ms]. 
A noticeable observation is the occasional occurrence of peaks that are higher than the typical height of modulations in the cloud. 
We interpret these peaks as structures where adjacent solitons have constructively interfered – an effect that is familiar from the collision of bright solitons~\cite{doi:10.1126/science.aal3220}, in which the apparent number and height of peaks can vary through interactions.

\begin{figure}[t]
\centering
\includegraphics[width=\columnwidth]{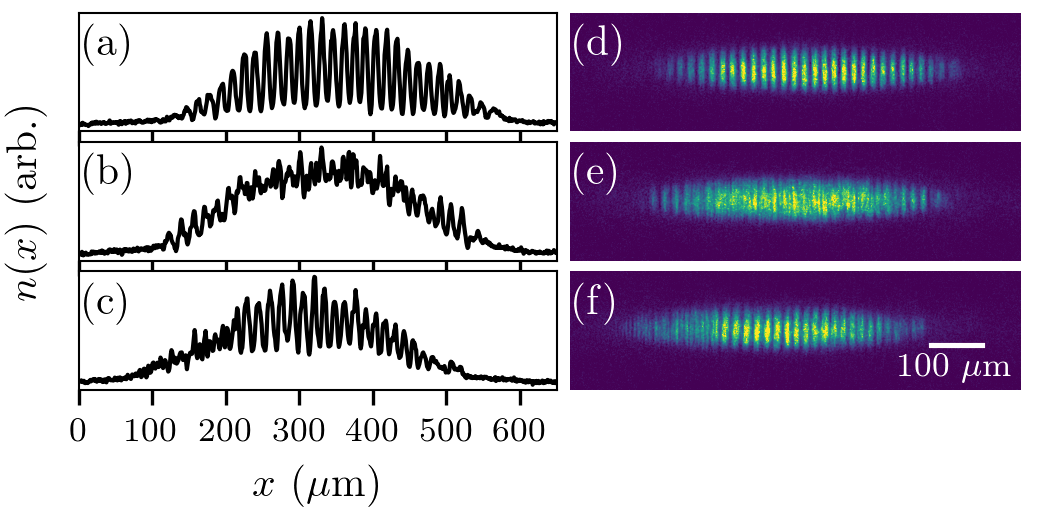}
\caption{Nucleation of a regular dense soliton train  featuring a transient fuzzy phase. 
Stage of reduced contrast and revival after 60~ms of winding. Evolution times after end of winding are (a,d) 70~ms, (b,e) 80~ms, (c,f) 90~ms. Panels (a-c) depict integrated cross sections of (d-f), respectively.}
\label{fig:60ms_fuzzyphase}
\end{figure}

For a winding time of 60 ms, a pattern periodicity of 16~$\mu$m is produced. 
This pattern periodicity is too fine to allow for the formation of pronounced plateaus that were observed in the previous case. 
Here, the pattern remains virtually unchanged for the first 70 ms, as dynamics gradually set in, shown in Figs.~\ref{fig:60ms_fuzzyphase}(a) and \ref{fig:60ms_fuzzyphase}(d). 
A prominent and reproducible feature of the evolution is the emergence of a fuzzy stage in which the contrast of the peak structure is  strongly diminished (90\% suppression of the specral pattern) in large areas of the BEC,  nearly disappearing into a uniform Thomas-Fermi profile [Figs.~\ref{fig:60ms_fuzzyphase}(b) and \ref{fig:60ms_fuzzyphase}(e)], followed by a revival of the peaks (to 62\% of the original spectral strength) [Figs.~\ref{fig:60ms_fuzzyphase}(c) and \ref{fig:60ms_fuzzyphase}(f)].
This peculiar stage is also reproduced qualitatively in our full 3D simulations when slight residual counterflow induces currents in the two components (see Supplementary Note 3).
We interpret this stage as a nearly simultaneous collision among the nonlinear waves throughout the array, enabled by the highly ordered initial conditions of the system and the good matching between the array periodicity and the natural length scale of the solitonic features.

\begin{figure}[b]
    \centering
    \includegraphics[width=\columnwidth]{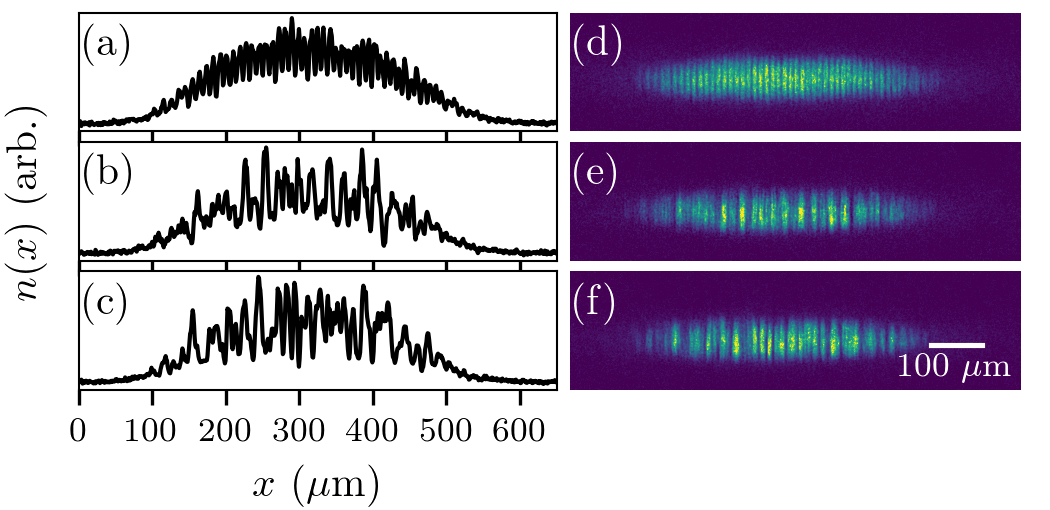}
    \caption{
    Experimental observations of dense collisional soliton complex emerging after 100~ms of winding. Evolution times after end of winding are (a, d) 0~ms, (b, e) 45~ms, and (c, f) 400~ms. Panels (a-c) show integrated cross sections of (d-f), respectively.}
    \label{fig:100ms_complexes}
\end{figure}

Finally, as the culmination of the progression described above, we arrive at the case of a dense pattern obtained after a winding time of 100~ms (with a pattern spacing of about 9~$\mu$m) presented in Fig~\ref{fig:100ms_complexes}. 
Here we observe dense collisional soliton complexes that maintain their qualitative character for long times. 
Figs.~\ref{fig:100ms_complexes}(a) and \ref{fig:100ms_complexes}(d) show the initial magnetization pattern. 
The pattern contrast appears reduced due to imaging effects including imaging resolution, expansion of the atomic cloud during time of flight, and residual thermal fraction.
After approximately 25~ms of evolution, the pattern becomes irregular. 
However, comparing the density profile after an evolution of 45~ms [Figs.~\ref{fig:100ms_complexes}(b) and \ref{fig:100ms_complexes}(e)] with that after 400~ms [Figs.~\ref{fig:100ms_complexes}(c) and \ref{fig:100ms_complexes}(f)] reveals that the pattern remains qualitatively unchanged in its overall characteristics (e.g., typical feature widths, peak heights, cloud size etc.) for surprisingly long times.

\begin{figure}[t]
\centering
\includegraphics{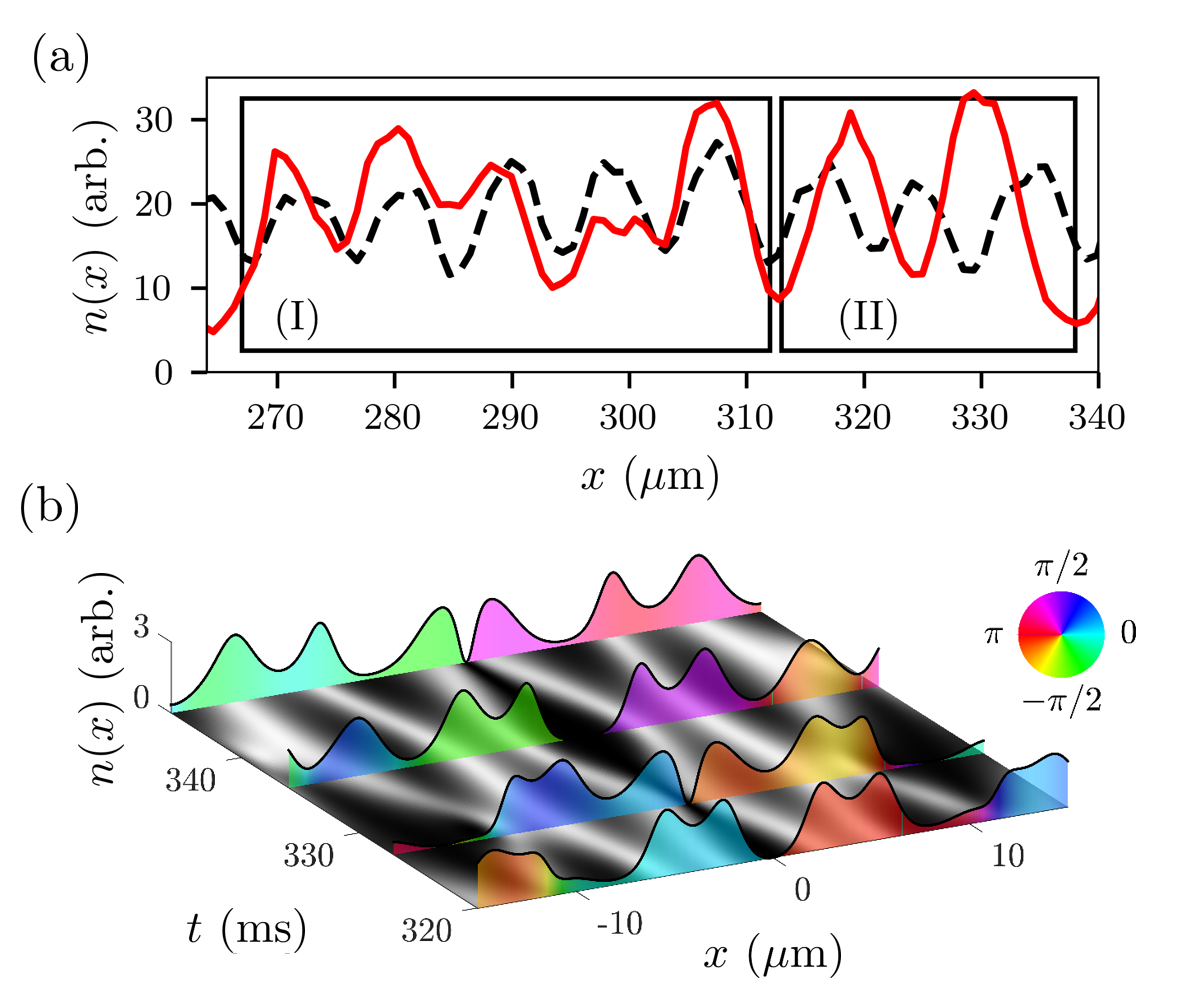}
\caption{Detailed observation of soliton crossings in dense collisional complexes emerging after 100~ms of winding. (a) Experimental cross sections after an evolution time of 150~ms (red, solid line) overlaid with initial cross section at 0~ms evolution (black, dashed) featuring regular soliton periodicity followed by tell-tale signatures of soliton collisions. 
(b) One-dimensional simulations show similar features in the time evolution of the wound system with the relative phase indicated by the color under the cross sections.} 
\label{fig:100ms_crosssections}
\end{figure}

A close inspection reveals that the disordered pattern maintains some structure: overlaying integrated cross sections with different evolution times (such as the characteristic example in Fig.~\ref{fig:100ms_crosssections} with a cross section obtained after 150~ms)  with the original pattern demonstrates that in many instances the original pattern periodicity is still maintained [Fig.~\ref{fig:100ms_crosssections}(a-I)] but the amplitude of the peaks has changed. 
A frequent observation is also that individual original peaks seem to be missing or strongly reduced and their neighboring peaks have grown in amplitude [Fig.~\ref{fig:100ms_crosssections}(a-II)]. 
Further corroborating evidence of this 
perspective is provided through one-dimensional simulations of the system provided in Fig.~\ref{fig:100ms_crosssections}(b), where the time evolution of the location and phase of individual solitons can be traced through time.

Such dynamics are familiar from colliding bright solitons and are intrinsic to a dense soliton gas. We also note in passing that it is interesting to compare, on a qualitative level, the cross sections in Figs.~\ref{fig:100ms_complexes}(e) and \ref{fig:100ms_complexes}(f) to those published in Ref.~\cite{Gelash2018}.
There, a 128-soliton solution was numerically calculated for the focusing, one-dimensional, nonlinear Schr\"{o}dinger equation -- a challenging task involving arbitrary-precision techniques to achieve reliable accuracy. 
What sets our observations apart is that the experimentally observed collisional complexes are maintained for long periods of time as the solitons are held together by the harmonic trap confining the BEC. 
This provides an experimental platform for the future experiments focusing on the intricate dynamics of interacting soliton complexes, including aspects of ``thermalization'' of the initial regular pattern. 

{\vspace{0.25cm}\noindent\large\textbf{Conclusion}}

Having reached the regime of dense collisional soliton complexes, our experiments provide a path to the study of dense soliton gases and soliton condensates within the framework of hydrodynamics in ultracold quantum gases. 
Recently, two pioneering experiments have reported the formation of a soliton gas in a long water tank~\cite{PhysRevLett.122.214502,Suret2020}. 
In the realm of BECs, a unique aspect is the possibility to dynamically change an axial harmonic confinement of the gas to modify
the relevant ``stress''. 
An analysis of the types of configurations experimentally
realized herein involving the development of the inverse scattering transform for the two-component nonlinear Schr\"{o}dinger type problem~\cite{El2020}, while out of the scope of the present manuscript, appears to us to be fully within reach. 
Excitations evolving on top of a regular soliton background, in terms of hypersolitons~\cite{Ma2016} or topological breathers~\cite{Mao2021,Bertola2022}, are another exciting research direction that will provide connections to condensed-matter systems.

{\vspace{0.25cm}\noindent\large\textbf{Methods}}

{\noindent\textbf{Experimental Methods}}

Our experimental technique for the generation of dense nonlinear excitations is based on a two-pulse Ramsey sequence in the presence of a small magnetic gradient. 
We begin with an elongated BEC of approximately $9 \times 10^5$ $^{87}$Rb atoms in the $\lvert F, m_F\rangle=\lvert1,-1\rangle$ hyperfine state held in an optical trap with harmonic trap frequencies of $\bm{\omega}/2\pi = \{2.5, 245, 258\}$ Hz.
In the presence of a 10~G magnetic bias field oriented in the vertical direction, a fast microwave $\pi/2$ pulse is applied which coherently creates an equal superposition of atoms in the $\lvert 1, -1\rangle$ and the $\lvert 2, -2\rangle$ states. 
This is followed by a wait time, referred to as the winding time $\tau$, during which the 10 G bias field is supplemented by a slight gradient of 5.12(1)~mG/cm along the long axis of the BEC. 
This leads to an accumulation of phase difference between the spin components which varies linearly across the length of the cloud.
Then a second $\pi/2$ pulse is applied which, depending on the phase between the precessing spins and the microwave pulse, transfers atoms into the $\lvert 2, -2\rangle$ state or back into the $\lvert 1, -1\rangle$ state.
The intra- and intercomponent scattering lengths in the system are all very similar, but not exactly equal (see Supplementary Note 2), which contributes to weakly miscible longtime dynamics. 
In all cases, the magnetic field is small and away from any Feshbach resonances, allowing the scattering lengths to be modeled with their zero field values.

This process produces a sinusoidal magnetization pattern which, crucially for the subsequent dynamics, contains domains with $\pi$ phase differences as shown in Fig.~\ref{fig:winding_intro}(b).
Notice that contrary to earlier works, e.g. Ref.~\cite{matthews}, there is no continuous application of the Rabi drive as we are interested in the undriven dynamics of the system after an initial pattern has been established.
Subsequent mean-field dynamics allow the sinusoidal phase pattern to relax into an array of alternating nonlinear waves.
The spatial periodicity of the produced pattern can be experimentally adjusted over a wide range by varying the winding time between the two microwave pulses.

Finally, we directly image the atomic density profile through absorption imaging of the $\lvert 2,-2\rangle$ spin state. 
The two spin components add together to form a nearly unmodulated Thomas-Fermi distribution because the energy associated with deforming the overall density of the cloud far exceeds the energy scale for spin-mixing in this system. 
The second spin state forms a complementary pattern to the observed state (see Fig.~\ref{fig:winding_intro}(a)) and therefore is omitted in the reported figures.

{\noindent\textbf{Numerical Methods}}

To simulate the dynamics of the nonlinear phenomena of interest, we utilize the following system of coupled 3D Gross-Pitaevskii equations~\cite{pethick,stringari}: 
\begin{equation}
\label{eq:3DCGPE}
    i\hbar \partial_t\Psi_j = \left(-\frac{\hbar^2}{2m} \nabla^2 + V(\mathbf{r})+\sum^2_{k=1} g_{jk} |\Psi_k|^2 \right) \Psi_j. 
\end{equation}
Here, $g_{jk}=4 \pi \hbar^2 a_{jk} N_j/m$ with $j=1,2$ indexing the relevant spin states,  $N_j=4.5 \times 10^5$ is the particle number per component, $a_{jk}$ are the 3D intra- ($j=k$) and inter-component ($j\neq k$) scattering lengths, and $m$ is the mass of a $^{87}$Rb atom. 
Additionally, the trapping potential $V(\mathbf{ r})=\sum_{\xi=x,y,z}m\omega_\xi^2 \xi^2/2$ is characterized by the aforementioned experimental trapping frequencies whose aspect ratio leads to a cigar-shaped geometry and hence precludes transverse instabilities~\cite{siambook,becbook2}. 

To initialize the dynamics, we first obtain the ground state of the self-interacting, yet intercomponent-decoupled, time-independent system of Eqs.~(\ref{eq:3DCGPE}).
Then, we imprint the desired complementary configuration between the two components with wave number $k_0$ and a  phase jump of $\pi$ between adjacent domains (see Supplementary Note 2) as shown in Fig.~\ref{fig:winding_intro}(b), while switching on the intercomponent coupling. 
Subsequently, the resulting waveform is evolved in time. 

{\vspace{0.25cm}\noindent\large\textbf{Data Availability}}
The data associated with this work are available from the corresponding author upon reasonable request.

{\vspace{0.25cm}\noindent\large\textbf{Code Availability}}
The code associated with this work are available from the corresponding author upon reasonable request.

{\vspace{0.25cm}\noindent\large\textbf{Author Contributions}}
S.M., T.B., and P.E. contributed to acquiring the experimental data and performing the analysis. G.K., S.I.M., and A.R. produced numerical simulation data and associated analysis. Funding for this work was acquired by P.G.K., P.S., and P.E. All authors contributed ideas and worked on preparing the manuscript.

{\vspace{0.25cm}\noindent\large\textbf{Author Competing Interests}}
The authors declare no competing interests.

{\vspace{0.25cm}\noindent\large\textbf{Acknowledgements}}

We acknowledge Maren Mossman for early contributions to the experimental investigation as well as Gino Biondini and Barbara Prinari for helpful theory discussions.
We acknowledge fruitful discussions lending context to this work at the Dispersive Hydrodynamics Program (2022) hosted by the Issac Newton Institute for Mathematical Sciences.
S.M, T.B, and P.E. acknowledge funding from NSF through Grant No. PHY-2207588. P.E. acknowledges support from the
Ralph G. Yount Distinguished Professorship at WSU.
S.I.M. acknowledges support from the NSF through a grant for ITAMP at Harvard University. 
This research was also supported in part by the National Science Foundation under Grant No. NSF PHY-1748958 (S.I.M and P.S.).
This work (P.S.) has been funded by the Deutsche Forschungsgemeinschaft (DFG, German Research Foundation) – SFB-925 – project 170620586. 
This material is based upon work supported by the US National Science Foundation under Grants DMS-2204702 and PHY-2110030 (P.G.K.).

\newpage
%\pagebreak
\clearpage
\widetext

\begin{center}
\large{\bf{Supplemental Material:
{\textquotedblleft}Observation of dense collisional soliton complexes in a two-component Bose-Einstein condensate{\textquotedblright} }}
\end{center}

\renewcommand{\thefigure}{S\arabic{figure}}
\renewcommand{\theequation}{S\arabic{equation}}
\setcounter{figure}{0}
\setcounter{equation}{0}
\twocolumngrid

\section{Supplementary Note 1: Experiment characterization}

\begin{figure}[b]
    \centering
    \includegraphics[width=\columnwidth]{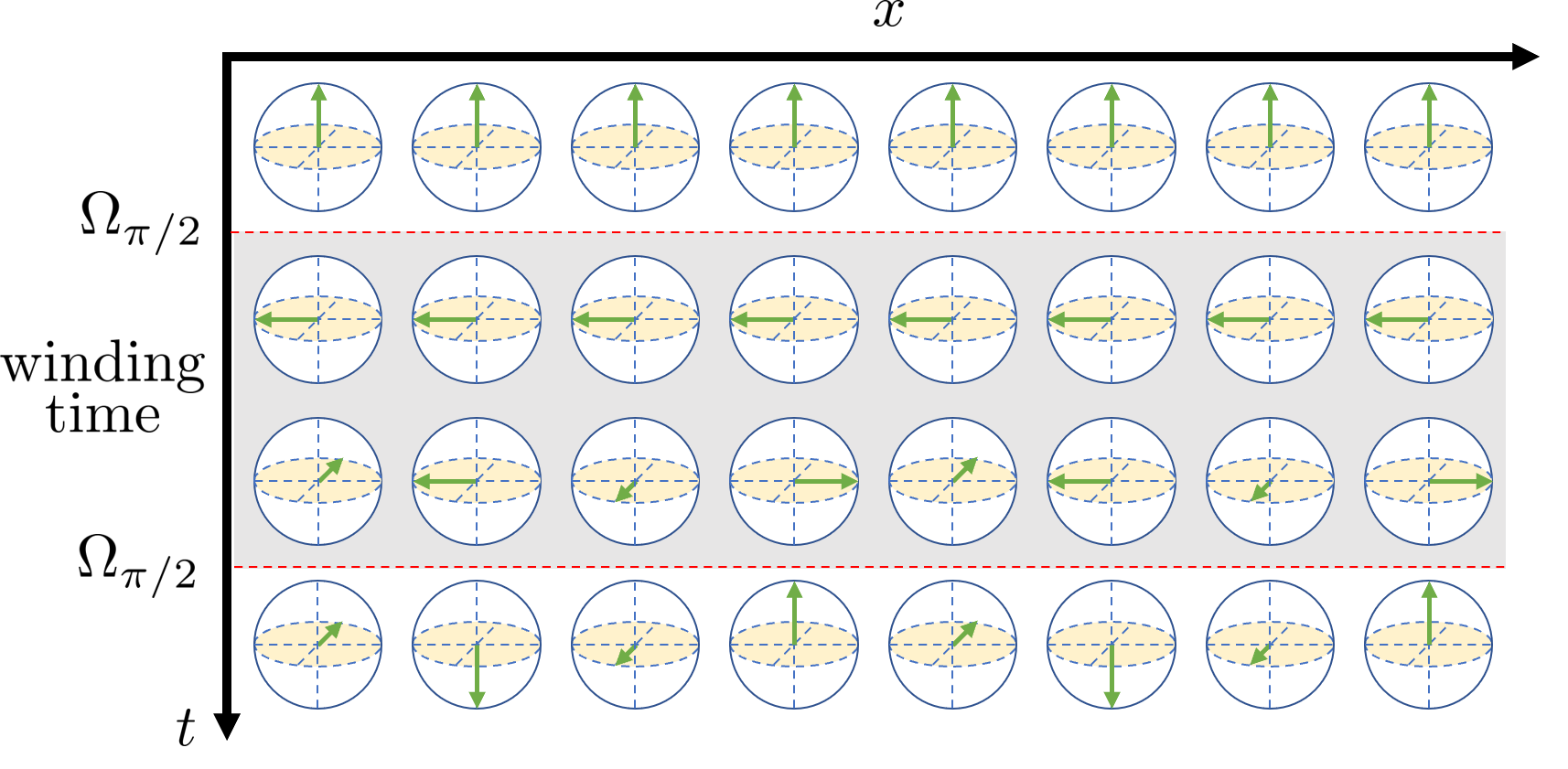}
    \caption{Schematic of the winding process where position in the BEC is represented on the horizontal axis and time progresses downwards. 
    A microwave pulse rotates the Bloch vector into the equatorial plane which then precesses at an increasing rate from left to right due to a magnetic field gradient during the winding time, shown in grey. A second microwave pulse then rotates the Bloch vectors again depending on their orientation at the end of the winding time.}
    \label{fig:exp procedure}
\end{figure}

In this section, we provide a more detailed description of the winding process described in the main text.
We begin with an elongated BEC entirely in the $\lvert 1,-1\rangle$ state, which we will refer to as spin up, with a magnetic bias field of 10~G in the vertical direction, perpendicular to the long axis of the condensate, along with a small magnetic field gradient along the long axis of the condensate. 
Here, it is helpful to consider a collection of Bloch spheres, one for each point along the long axis, as schematically illustrated in Fig.~\ref{fig:exp procedure}, with all Bloch vectors pointing up and in phase.

A near-resonant microwave pulse is tuned to coherently transfer half of the atomic population to the $\lvert 2,-2\rangle$ state, which we will refer to as spin down, with a pulse time of approximately 0.1~ms. 
The pulse produces a uniform spin mixture across the condensate.
Equivalently, this corresponds to a $\pi/2$ rotation of the Bloch vectors, leaving them in phase along the equator of the Bloch sphere.
Hence, we refer to this pulse protocol as a $\pi/2$ pulse.
The superposition state, shown in Fig.~\ref{fig:exp procedure} on the second row, is then allowed to evolve in the optical trap for a variable amount of time, referred to as the winding time $\tau$.

During the winding time, each Bloch vector precesses at a different rate and adjacent Bloch vectors will acquire a phase difference proportional to their energy difference as $\delta\phi=(\delta U/\hbar) t$.
In the single particle limit, this energy difference is given by the differential Zeeman splitting induced by the small gradient of the magnetic field strength along the long axis of the BEC.
We take the energy to be locally linear in the magnetic field for small changes giving $\delta\phi = \left(\frac{1}{\hbar}\frac{\partial U}{\partial B}\right)_{B=B_0}\Delta B\ t$, where the variation in the energy between the $\lvert 1,-1\rangle$ and $\lvert 2,-2\rangle$ states of $^{87}$Rb at $B_0=10$~G is $2.0972h$ MHz/G using the Breit-Rabi formalism~\cite{steck01.01}.
The change in magnetic field between two points in the condensate can then be 
approximated by assuming that the gradient is linear along the long axis of the condensate, leading to 
\begin{equation}
    \delta\phi = \left(\frac{1}{\hbar}\frac{\partial U}{\partial B}\right)_{B=B_0}\left(\frac{\partial B}{\partial x}\right)\Delta x\ t.
\label{eq:wingingphase}
\end{equation}

\begin{figure}[t]
\centering
\includegraphics[width=\columnwidth]{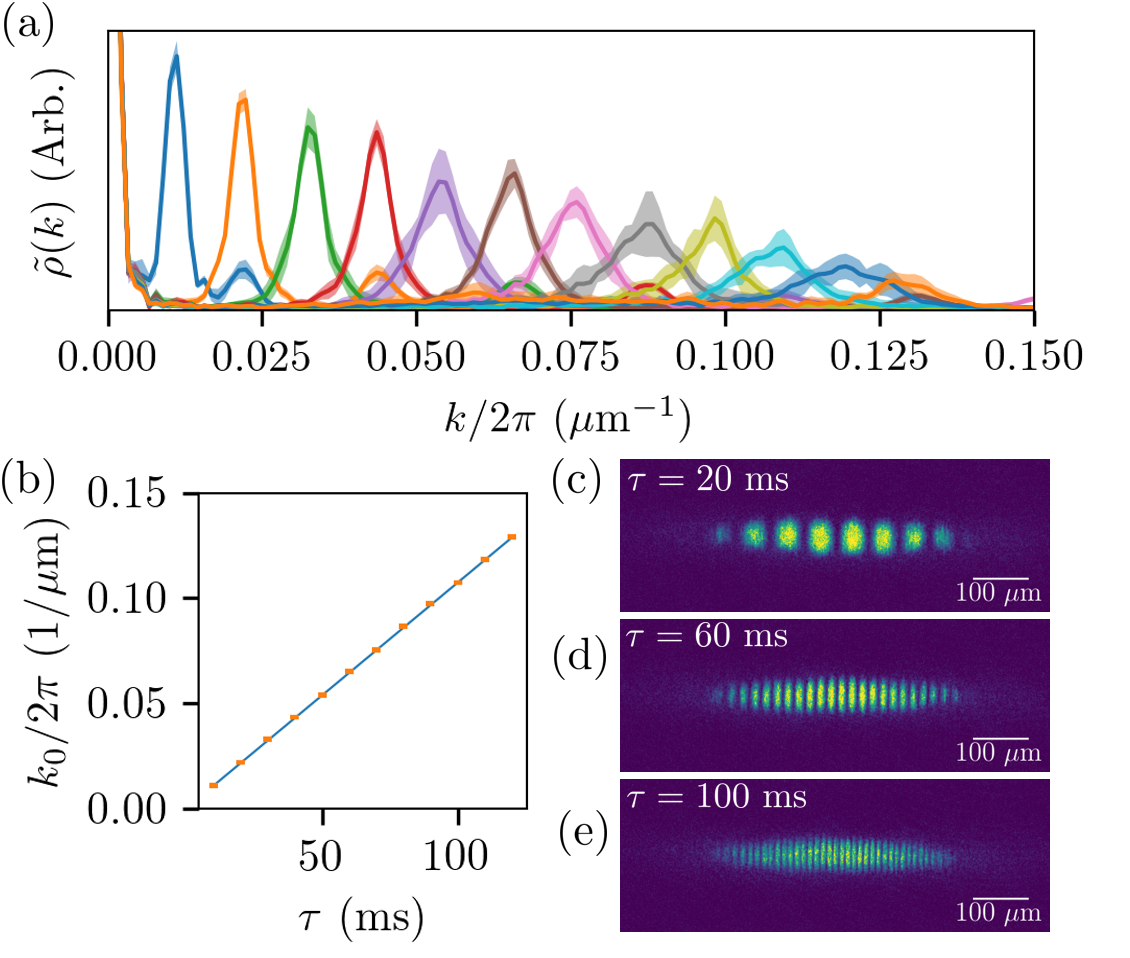}
\caption{(a) Experimental Fourier spectra for a range of winding times from $\tau=10$~ms to $\tau=120$ ms in steps of 10 ms. 
The $k$-vector of the  corresponding fundamental peak $k_0$ increases monotonically with $\tau$. 
Each spectrum is averaged over ten experimental realizations for a given winding time. 
The standard deviation in the averaged spectra is shown as a shaded region around each curve. 
(b) The center value wavenumber from fits to the peaks in panel (a). 
The standard error in each fit is smaller than the size of the markers. 
A linear fit indicates a winding rate of $1.074(2)\ (\mu\mathrm{m\ s})^{-1}$. (c)--(e) Absorption images of the wound configuration after $\tau=20$, 60 and 100~ms of winding, respectively.}
\label{fig:winding_up}
\end{figure}

Finally, a second $\pi/2$ pulse is applied which continues to rotate the Bloch vector about the same axis as the initial pulse.
The last row of Fig.~\ref{fig:exp procedure} shows how, depending on the phase acquired between pulses, the Bloch vectors will be unaffected, rotated back to the up state, rotated to the down state, or someplace in between along a sinusoidal pattern.
When the phase difference between two points in the condensate is $2\pi$, both of those regions will be rotated in the same way by the final $\pi/2$ pulse and thus represent a full wavelength of the final magnetization pattern.
Fig.~\ref{fig:winding_up}(a) presents the progression of the winding density as the winding time is increased obtained through a Fourier transform of the density. 
Choosing $\Delta x=\lambda$, we identify the winding wavenumber 
\begin{equation}
    k_0(t)=\frac{2\pi}{\lambda}=\left(\frac{1}{\hbar}\frac{\partial U}{\partial B}\right)_{B=B_0}\left(\frac{\partial B}{\partial x}\right)t. \label{eq:koft}
\end{equation}
The slope extracted from Fig.~\ref{fig:winding_up}(b) is the winding rate in Eq.~(\ref{eq:koft}) from which we determine the magnetic field gradient in our experiment to be 5.12(1)~mG/cm. 
Examples of the wound configuration are shown for winding time $\tau=20$, 60, and 100 ms in Fig.~\ref{fig:winding_up}(c)-(e), respectively.

As each spin domain of the prepared magnetization pattern has undergone an additional phase rotation on the Bloch sphere relative to the adjacent domains, we would expect the condensate wave function to also acquire phase gradients of $\pi$ across each winding.
In the condensate picture, the effect of the small magnetic gradient during the winding time can be understood as a force which accelerates the spin-up and spin-down components in opposite directions. 
This acceleration results in a relative velocity corresponding to the phase difference in the condensate wave function between the two spin components. 
This phase difference then emerges as phase windings after the final $\pi/2$ pulse remixes the two spin components, which then localize as phase jumps under mean field interactions to produce nearly stationary dark-bright solitons.

\section{Supplementary Note 2: Mean-field implementation of the soliton arrays}

To numerically study the nonequilibrium dynamics of the distinct arrays of nonlinear excitations we employ the dimensionless version of the 3D coupled Gross-Pitaevskii equations~\cite{Bersano2018,Katsimiga2020} provided in the main text, see also Eq.~(1). 
As such, the underlying particle ($N_1=N_2=N/2=4.5 
\times 10^5$) and mass ($m_1=m_2=m$) balanced 
$^{87}$Rb mixture is described by
\begin{equation}
\label{eq:3DGPE}
\begin{split}
    i &\partial_t \Phi_1(\mathbf{r},t) = -\frac{1}{2} \nabla^2 \Phi_1(\mathbf{r},t) + V({\bf r}) \Phi_1(\mathbf{r},t) \\&+ (4 \pi a_{1,1} N_1 |\Phi_1(\mathbf{r},t)|^2 + 4 \pi a_{1,2} N_2 |\Phi_2(\mathbf{r},t)|^2) \Phi_1(\mathbf{r},t),
    \\
    i &\partial_t \Phi_2(\mathbf{r},t) = -\frac{1}{2} \nabla^2 \Phi_2(\mathbf{r},t) + V({\bf r}) \Phi_2(\mathbf{r},t) \\&+ (4 \pi a_{2,1} N_1 |\Phi_1(\mathbf{r},t)|^2 + 4 \pi a_{2,2} N_2 |\Phi_2(\mathbf{r},t)|^2) \Phi_2(\mathbf{r},t),
\end{split}
\end{equation}
Here, the Laplacian operator is $\nabla^2 \equiv 
\partial^2_x + \partial^2_y + \partial^2_z $, whilst the employed 3D scattering lengths for $^{87}$Rb are $a_{1,1} = 100.40(10)$, $a_{2,2} = 98.98(4)$, and $a_{1,2} = 98.98(4)$ in units of the Bohr radius $a_0$~\cite{scattering_lengths}.
State 1 is $\lvert F, m_F \rangle = \lvert 1,-1\rangle$, state 2 is $\lvert 2,-2\rangle$, and $m$ is the mass of $^{87}$Rb.
The dimensionless 3D parabolic external potential reads $V({\bf r})=\frac{1}{2}\left( x^2+\left(\omega_y/\omega_x\right)^2 y^2+\left(\omega_z/\omega_x\right)^2 z^2 \right)$ with ${\bf r}=(x,y,z)$. 
Following the experimental implementation the axial and transverse trapping frequencies are $\left(\omega_x,\omega_y,\omega_z\right)=2\pi \times \left(2.5, 245, 258\right)$~Hz justifying a highly-elongated (cigar-shaped) geometry possessing an aspect ratio $\omega_x/\omega_y\approx \omega_x/\omega_z \approx 0.01$. 

Moreover, the rescaling used for the spatial and temporal coordinates is  
$x'=a^{-1}_{\rm {ho}}x$, $y'=a^{-1}_{\rm {ho}}y$, $z'=a^{-1}_{\rm {ho}}z$, with $a_{\rm {ho}}=\sqrt{\hbar/m\omega_x}$ denoting the harmonic oscillator length along the longitudinal $x$-direction, and $t'=\omega_x t$, respectively. 
Accordingly, the wave function of each hyperfine state ($j=1,2$) is rescaled as $\Phi_j(x',y',z')=\sqrt{N_j/a^3_{\rm {ho}}}\Psi_j(x,y,z)$ and the Laplacian is rescaled as $\nabla^2_{r'}=a^2_\mathrm{ho}\nabla^2$. 
Note that in Eq.~\eqref{eq:3DGPE} we dropped the primes for clarity.

To emulate the experimental preparation we first obtain the ground state of the intracomponent interacting 3D binary system,  
setting $a_{12}=0$ and utilizing a fixed point iterative scheme of the Newton type~\cite{kelley2003solving}. 
The total density of the system features
a Thomas-Fermi profile.
As a subsequent step, we craft on top of the aforementioned ground states the following sinusoidal ansatz [see Fig.~1(a) in the main text]
\begin{eqnarray}
\label{crafting1}
\tilde{\Phi}_1(\mathbf{r})&=&\sqrt{\cos^2\left(k x\right)}\Phi_1\left(\mathbf{r};0\right)
e^{i\pi \cos \left(k x/2\right)},\\
\label{crafting2}
\tilde{\Phi}_2(\mathbf{r})&=&\sqrt{\sin^2\left(k x\right)}\Phi_2\left(\mathbf{r};0\right)
e^{i\pi \cos \left(\pi/4+k x/2\right)}, \nonumber \\ 
\end{eqnarray}
where $\Phi_j(\mathbf{r};0)$ is the ground state of each component at $t=0$. 
Notice that these crafted wave functions of the two 
components along the longitudinal $x$-direction are complementary with respect to one another. 
An almost perfect total Thomas-Fermi density profile occurs in the decoupled case with weak spatial undulations appearing for increasing (principal) wavenumbers $k=k_0/(2 \pi)$.
Finite intercomponent interactions for a fixed $k_0$ lead to more pronounced spatial undulations. 
To trigger the dynamics we switch on the intercomponent interactions and let the above system evolve for times up to $t=200$~ms. 
The spatiotemporal evolution of the two-component bosonic system is captured using a fourth-order (in time) Runge-Kutta method characterized by temporal and spatial discretization $dt=10^{-4}$ and $dx=0.03$, $dy=dz=0.05$ respectively. 
Also, a second order finite difference scheme is employed to resolve the spatial derivatives.

\begin{figure}%[h]
\centering
\includegraphics[width=\linewidth]{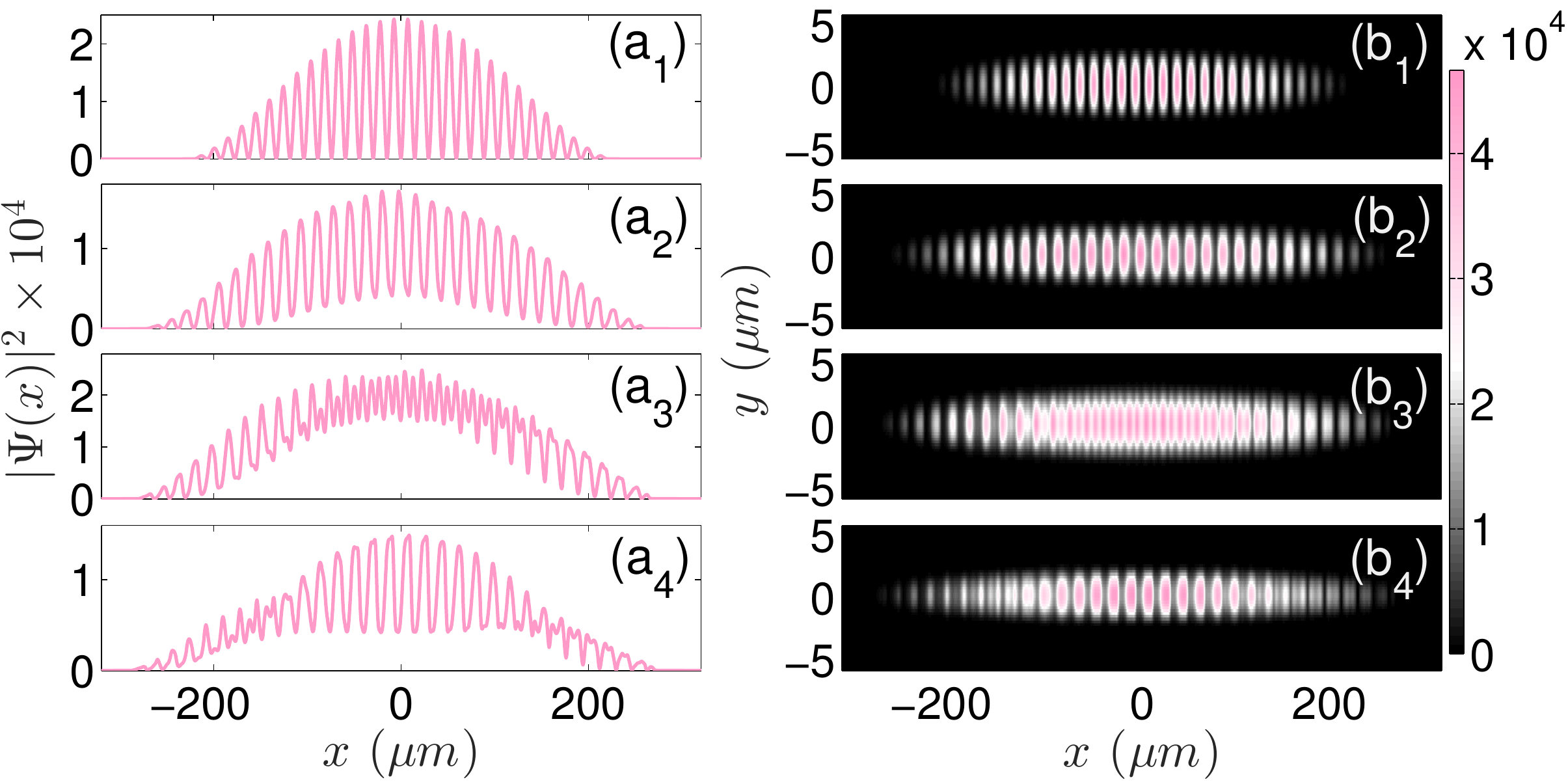}
\caption{Density profiles with the $k_0$ selected to match the $\tau=60$~ms windings integrated along (${\rm a1}$)--(${\rm a4}$) $y$-$z$ plane and (${\rm b1}$)--(${\rm b4}$) $z$-direction. The times depicted from top ($a_1$), ($b_1$) to bottom ($a_4$), ($b_4$) are $t=0,70, 80$, and $90$~ms.}
\label{fig:density_60wind}
\end{figure}

\section{Supplementary Note 3: Dynamical features of the distinct soliton arrays}

\begin{figure}%[h]
\centering
\includegraphics[width=\linewidth]{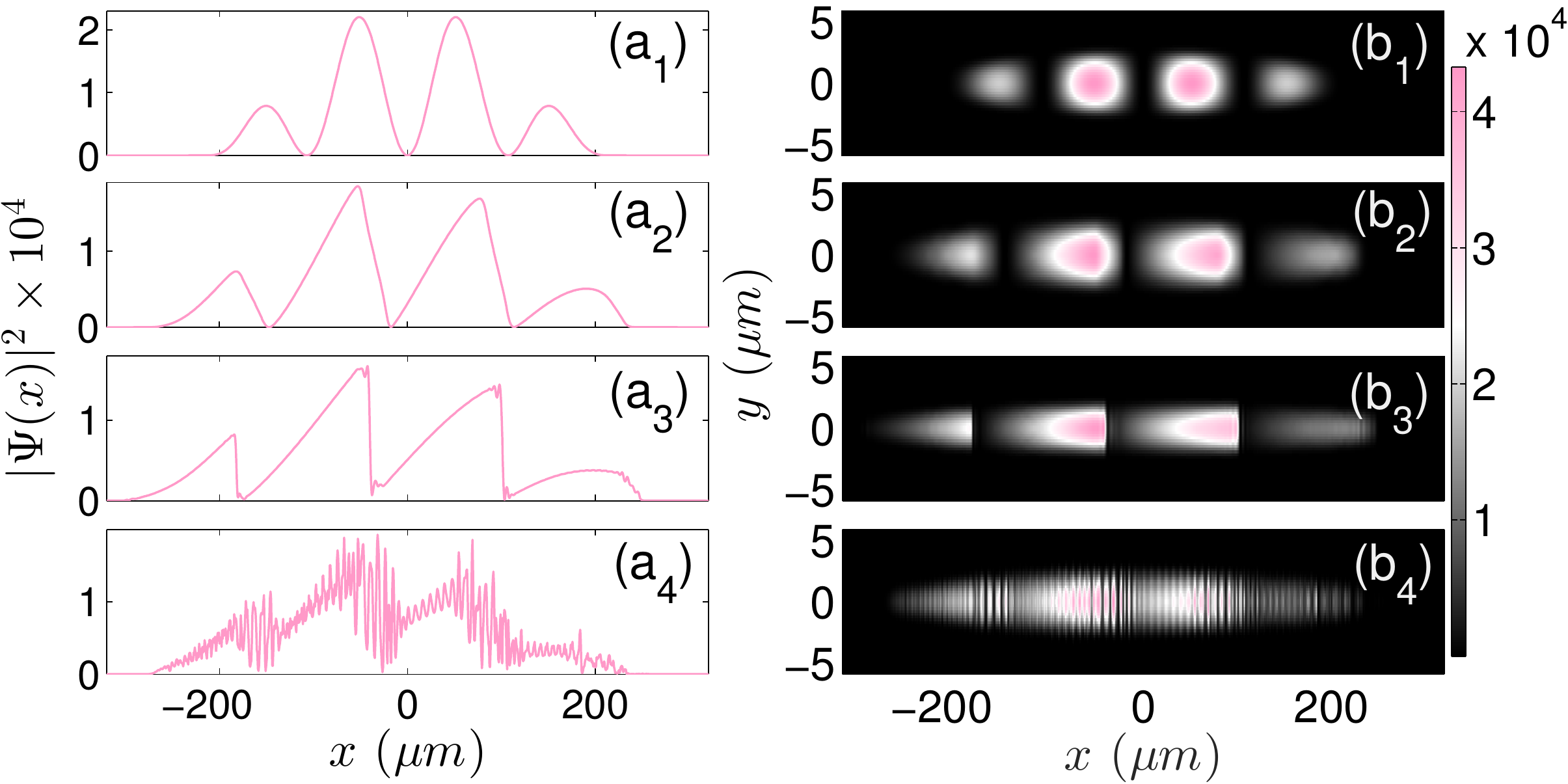}
\caption{Density snapshots using the $k_0$ that matches $\tau=10$~ms windings depicting the integrated profiles along (${\rm a1}$)--(${\rm a4}$) y-z plane and (${\rm b1}$)--(${\rm b4}$) z direction. The time-instants illustrated from top ($a_1$), ($b_1$) to bottom ($a_4$), ($b_4$) correspond to $t=0,80, 125$, and $190$~ms. In all cases, only the second component is visualized since the first is complementary to it. 
A counterdirectional displacement of the component's harmonic trap enforces intercomponent counterflow with dark solitons initially emerging at the right shifted steep edges of the cloud.}
\label{fig:density_10wind}
\end{figure}

\begin{figure}%[h]
\centering
\includegraphics[width=\linewidth]{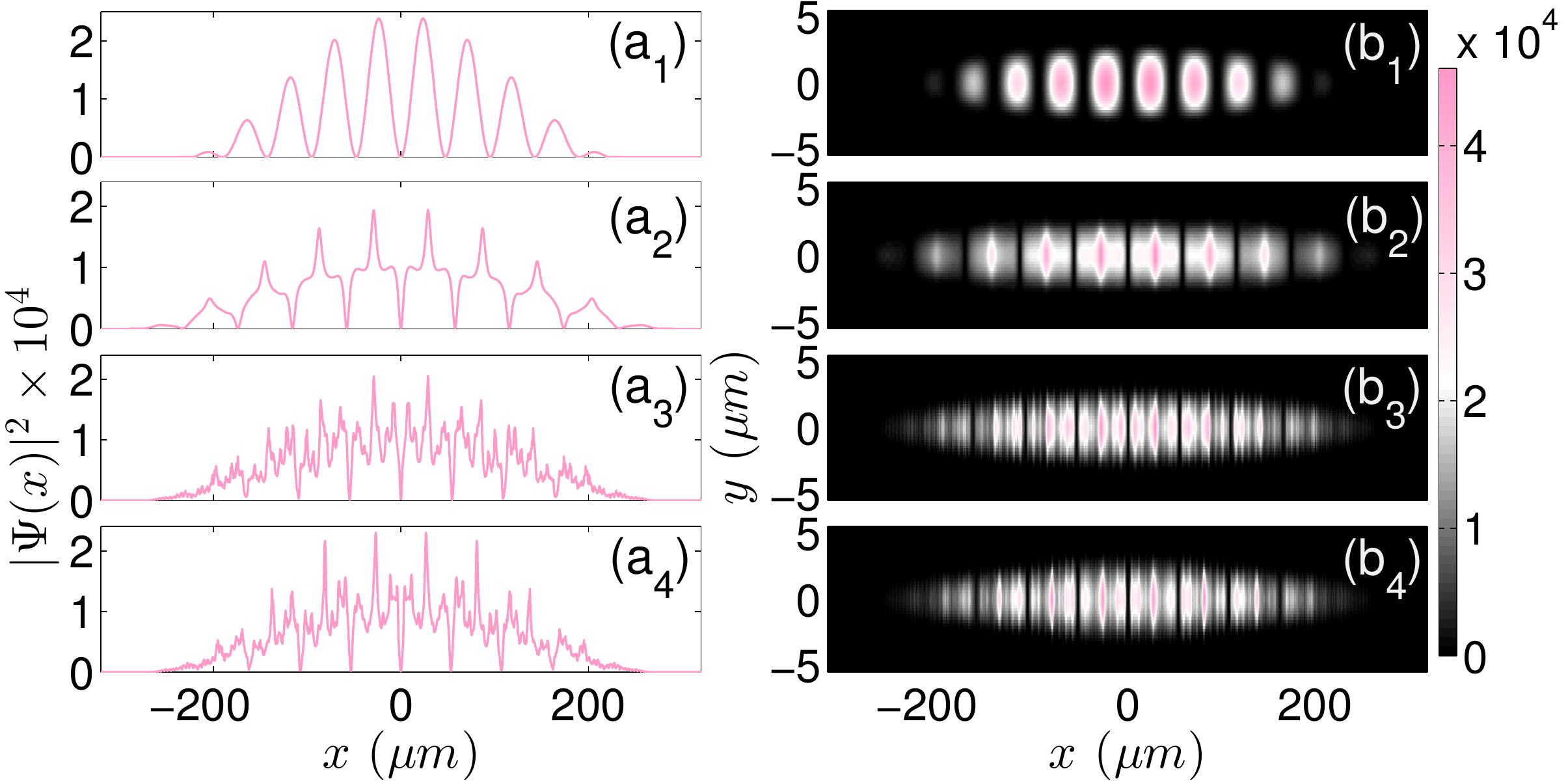}
\caption{Density profiles, with the relevant $k_0$ matching the  $\tau=20$ ms windings, integrated along (${\rm a1}$)--(${\rm a4}$) y-z plane and (${\rm b1}$)--(${\rm b4}$) z direction. The evolution times shown from top ($a_1$), ($b_1$) to bottom ($a_4$), ($b_4$) are $t=0,80, 140$, and $190$~ms. Solely the second component is illustrated since the first is complementary to it. Progressively antidark structures develop throughout the cloud.
}
\label{fig:density_20wind}
\end{figure}

Characteristic density profiles during the nonequilibrium dynamics of the 3D system of coupled GPEs [Eq.~(\ref{eq:3DGPE})] are presented in Figs.~\ref{fig:density_10wind}, \ref{fig:density_20wind} and \ref{fig:density_60wind} for three winding protocols discussed in the main text, namely for $\tau = $ 10~ms, 20~ms, and 60~ms, allowing us to dynamically generate a plethora of nonlinear excitations. 

We start with the dilute winding pattern of $\tau=10$~ms while also considering a counterflow between the two components. 
Counterflow is triggered via a counterdirectional displacement of the individual component's harmonic trap with strength 3~${\rm \mu m}$ and 6.5~${\rm \mu m}$ following the experimental procedure. 
The initial regular pattern consisting of four (five) density peaks in the second (first) component [Figs.~\ref{fig:density_10wind}(a1) and \ref{fig:density_10wind}(b1)] gradually develops a triangular sharp edged configuration [Figs.~\ref{fig:density_10wind}(a2) and \ref{fig:density_10wind}(b2)]. 
This is attributed to the counterflow featured by the participating hyperfine states. 
We remark that in the absence of counterflow, dark-antidark solitons build upon the components in a complementary fashion [Fig.~\ref{fig:density_20wind}]. 
These persistent waveforms also emerge for larger winding protocols, see our discussion below and the main text.    
Subsequently, these sharp density fronts develop dark solitons at their edges [Figs.~\ref{fig:density_10wind}(a3) and \ref{fig:density_10wind}(b3)] stemming from the interference among the components. 
These solitary waves progressively spread throughout the bosonic cloud [see in particular Figs.~\ref{fig:density_10wind}(a4) and \ref{fig:density_10wind}(b4)]. 
Notice that all the above-described qualitative features are in line with the experimental observations [see Fig.~2 in the main text]. 

\begin{figure*}[t]
    \includegraphics[]{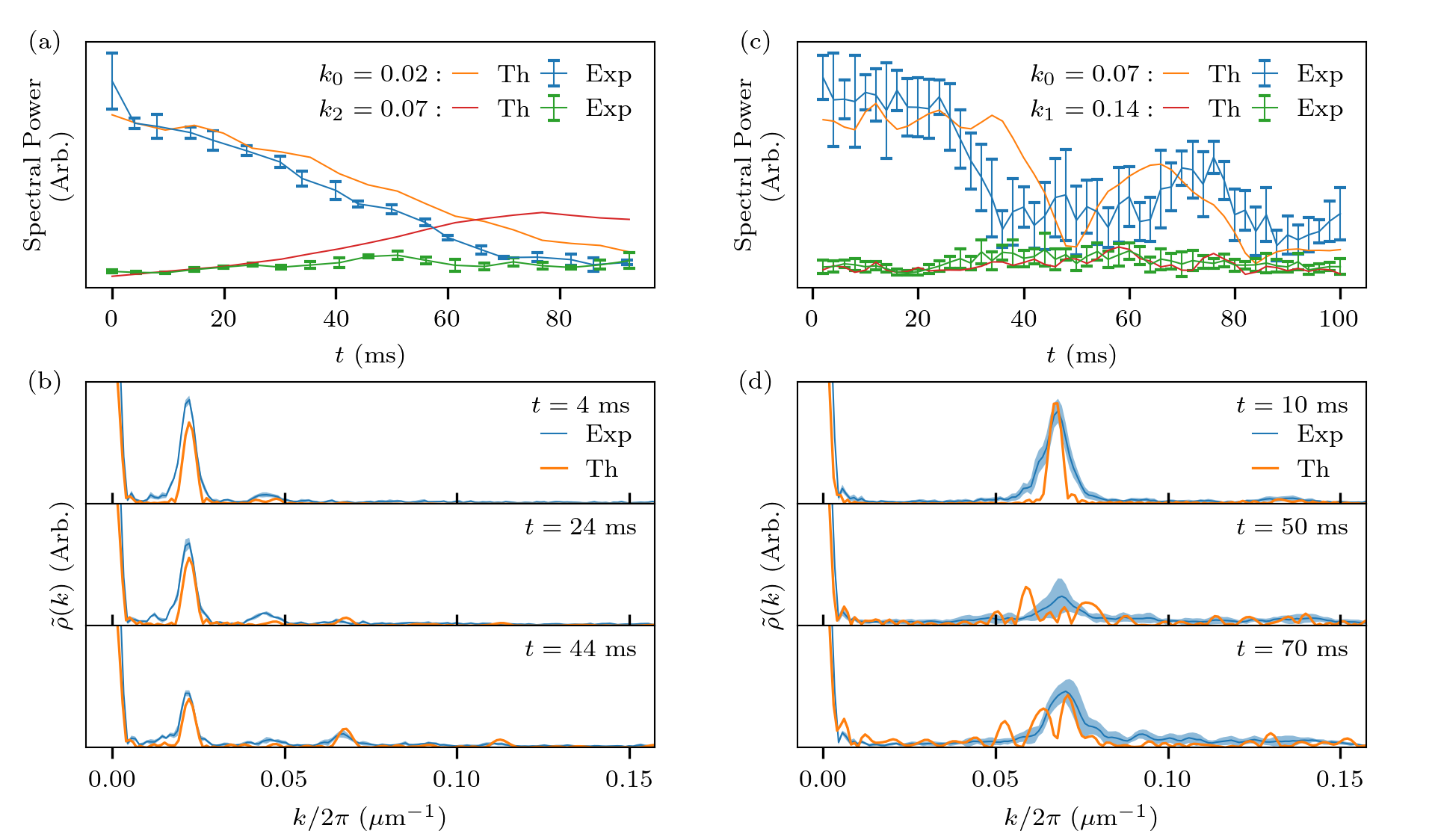}
    \caption{Time evolution of spectral features of soliton arrays.  Winding times of $\tau = 20$~ms (a-b) and $\tau = 60$~ms (c-d) are featured. Panels (a, c) show the evolution of the relative spectral power associated with the two most dominant spectral peaks resulting from the Fourier transform of the density. Panels (b, d) show example spectra at various evolution times for the two winding cases. Experimental error bars are given by standard deviation over independent realizations.}
    \label{fig:fftsummary}
\end{figure*}

To investigate the dynamics of denser soliton trains we employ a $\tau=20$~ms winding pattern in the absence of counterflow. 
At initial times a periodic pattern illustrated in Figs.~\ref{fig:density_20wind}(a1) and \ref{fig:density_20wind}(b1) occurs, acquiring a triangular shape around $t=35$~ms. 
The latter subsequently deforms into an antidark configuration [Figs.~\ref{fig:density_20wind}(a2)-(a4) and \ref{fig:density_20wind}(b2)-(b4)] as it is also demonstrated in the experimental cross section, e.g., of Fig.~3(a) and 3(d) of the main text. 
These antidark structures persist in the course of the evolution even when drastic dynamics sets in with the antidarks becoming more irregular in amplitude [Figs.~\ref{fig:density_20wind}(a3) and \ref{fig:density_20wind}(b3)]. 
Additionally, at later times ($t>100$~ms) due to adjacent soliton interactions high density peaks appear in both components having a complementary nature [Figs.~\ref{fig:density_20wind}(a4) and  \ref{fig:density_20wind}(b4)]. 
Notice that the amplitude of these density humps is significantly larger when compared to the overall background one. 
This observation is also captured by the actual experiment shown in Fig.~3(f) of the main text.

Next, we turn to the $\tau=60$~ms wound scenario.
To numerically investigate the influence of counterflow in this case, we apply a small amount of differential trap offset on the two spin components.
Fig.~\ref{fig:density_60wind}(a1-b1) show the regularity of the initial pattern which is sustained for evolution times up to approximately $20$~ms then followed by a loss and revivial of the contrast in the spin modulation. 
Such cycles in the depth of the waveforms take place e.g. in the vicinity of $35$~ms and around $80$~ms, an outcome that can be traced back to the presence of currents induced by the aforementioned counterflow. 
The second of these cycles is shown in Figs.~\ref{fig:density_60wind}(a2-b2) and \ref{fig:density_60wind}(a4-b4) for 70 ms, 80 ms, and 90 ms, respectively. 
This behavior parallels the experimental observations.

To facilitate a comparison of time scales between the experimental and numerical results, Fig.~\ref{fig:fftsummary} shows the time evolution of the relative spectral power for the dominant peaks responsible for the periodicity of the array. 
Experimental results and 3D numerics without counterflow are compared for the two cases of initial winding times of 20~ms (a-b) and 60~ms (c-d). 
Figs.~\ref{fig:fftsummary}~(a,c) show the time evolution of the relative strength of the primary spectral peak at wave number $k_0$ along with the most notable higher harmonic.
Figs.~\ref{fig:fftsummary}~(b,d) show example spectra from the marked time segments to provide information about the spectral properties of the soliton array.
Considering the phenomenological richness of the observed nonlinear dynamics and their sensitivity to initial conditions, the agreement between experimental and numerical time scales is reasonable. 
On a qualitative level, all key features of the experiment are well reproduced by the 3D mean-field simulations as described in the main text.

Additionally, a video depicting the real-time evolution after $\tau=100$~ms winding in a one-dimension simulation is provided in~\url{https://wolke.physnet.uni-hamburg.de/index.php/s/iyYkbzBb8pBWLzM}. 
These simulations are in qualitative agreement with the experimental observations made in Fig.~5 of the main text. 
Namely, the emergent highly dense soliton arrays feature collisions leading to spatial irregularities, while maintaining their overall characteristics for large evolution times.

\end{document}